\documentclass[11pt]{article}
\usepackage{moriond,epsfig}

\def\Journal#1#2#3#4{{#1} {\bf #2}, #3 (#4)}

\def\NPB{{\em Nucl. Phys.} B}

\def\PRD{{\em Phys. Rev.} D}

\def\PRE{\em Phys. Rev.}
\def\MPLA{{\em Mod. Phys. Lett.} A}

\def\be{\begin{equation}}
\def\ee{\end{equation}}
\def\bea{\begin{eqnarray}}
\def\eea{\end{eqnarray}}

\def\Nuno {\mathrm{\chi}_1^0}
\def\Ndue {\mathrm{\chi}_2^0}

\def \gluino {\tilde{\mathrm g}}
\def \sbottom {\tilde{\mathrm b}}
\def \squark {\tilde{\mathrm q}}
\def \slepton {\tilde{\ell}}

\def \stau {{\tilde{\tau}}}

\begin{document}
\vspace*{4cm}
\title{SPARTICLE RECONSTRUCTION AT LHC}

\author{ ALESSIA TRICOMI \\
on behalf of the ATLAS and CMS Collaborations}

\address{Department of Physics and Astronomy, University of Catania, and INFN Catania, \\ 
Via S. Sofia 64, I-95123 Catania, Italy}

\maketitle\abstracts{
In this report a review of recent studies made to understand the capability to 
discover and measure properties of SUSY particles at LHC is discussed. The expected 
resolution on sparticle masses is discussed on the basis of studies performed 
by the ATLAS and CMS collaborations.}

\section{Introduction}
\label{intro}

One of the main purposes of the LHC collider is to search for 
Physics beyond the Standard Model (SM). Supersymmetric (SUSY) extensions 
of SM~\cite{susy} predict the existence of superpatners for all ordinary particles.
If supersymmetry exists at the electroweak scale, it could
hardly escape detection at LHC. The centre-of-mass energy of 14 TeV, 
available at LHC, extends the searches for 
SUSY particles up to masses of 2.5 to 3 TeV/$c^2$. 
In R-parity conserving models, the stable Lightest Supersymmetric 
Particles (LSP), which escape 
detection, lead to events characterized by large missing energy.  
Usually, squarks and/or gluinos are the most abundant sparticles produced at LHC. 
They decay in a number of steps to quarks, charginos, neutralinos, 
sleptons, W, Z, Higgs bosons, etc. 
These events are expected to show up at LHC
via an excess of multijet+$E^{\mathrm {miss}}_{\mathrm T}$+multilepton final
states compared to the SM expectations~\cite{baer}. \\
A significant part of the efforts in preparation for 
the LHC startup is being spent in the simulations of the new physics 
potential. In the past years, the simulation studies have been mainly 
devoted to understand the discovery reach of SUSY particles through 
inclusive studies. In these studies, the typical discovery 
strategy consists in searching for an excess of events with a 
given topology with respect to the Standard Model expectations. 
A variety of final state signatures has been
considered. 
Inclusive studies have mainly been carried out in the framework
of mSUGRA~\cite{abdu,atlasph}, with five independent parameters: the common 
gaugino mass $m_{1/2}$, the common scalar mass $m_0$, the common 
trilinear scalar coupling $A_0$, the ratio of the vacuum expectation values 
of the two Higgs doublets $\tan\beta$ and the sign of the Higgsino mixing 
parameter $\mu$. Strong indications exist, however,  
that the overall SUSY reach in terms of masses of squarks and gluinos 
is very similar in most the R-parity conserving scenarios, provided that 
$m_{\Nuno}\ll m_{\gluino,\squark}$. This has been shown to be the case 
for the AMSB model~\cite{ab18} and even for the MSSM~\cite{ab19}.
 Already 
with only 1 fb$^{-1}$ of integrated luminosity, LHC should be able 
to discover squarks and gluinos if their masses do not exceed about 
1.3~TeV/$c^2$. 
With 100 fb$^{-1}$ the reach can be extended up to masses $m_{\squark}\sim
 m_{\gluino}\sim 2.5$ TeV/$c^2$. The entire plausible domain of EW-SUSY parameter space 
for most probable value of tan$\beta$ can be probed. \\
 Determining the masses of supersymmetric particles is more difficult,
because each SUSY event contains two LSP's, and there are not enough 
kinematic constraints to determine the momenta of these particles. The main goal
of this report, is to show the potential of the ATLAS~\cite{atlas} 
and CMS detectors~\cite{cms} to reconstruct SUSY particles  
and the achievable mass resolution. 

\section{SUSY Particle Measurements}
\label{susypart}

The reconstruction of sparticle masses and the determination of their properties might 
help in probing different models. In order to reconstruct sparticle masses, a different 
strategy with respect to that developed for inclusive analyses must  
be used. The typical procedure consists of several steps: (i) choose a set of benchmark 
points compatible with all existing measurements and spanning the whole 
SUSY parameter space, (ii) choose a particular decay chain, (iii) get the mass spectrum, 
exploit the kinematical 
properties such as the presence of characteristic end points or thresholds, 
and finally (iv) try to reconstruct sparticles using constraints from 
various mass combinations~\cite{allanach}. \\
The choice of benchmark points depends on the purpose of the
actual investigation. 
A large variety of benchmark points have been proposed in the past.  
The studies presented in this report rely on a set of points~\cite{atlasph,batt,sps}, 
which takes into account constraints from LEP results, cosmology measurements 
and low-energy experiments. 

\subsection{Dilepton edge reconstruction}
\label{subsec:dilep}

\begin{figure}[h]
\vfill \begin{minipage}{.49\linewidth}
\begin{center}
\mbox{\epsfig{figure=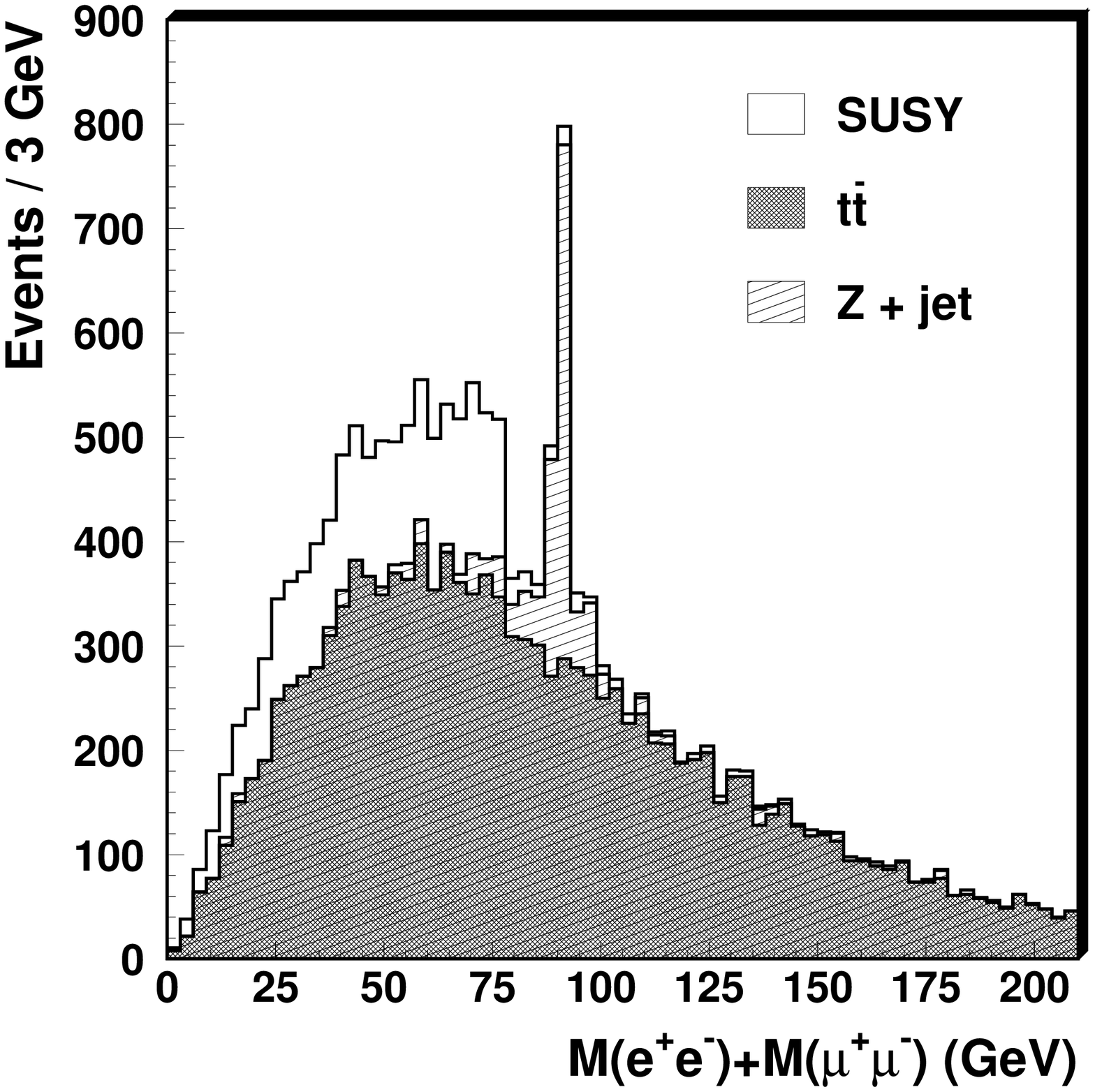,width=0.7\linewidth,clip=}}
\end{center}
\end{minipage}\hfill
\begin{minipage}{.49\linewidth}
\begin{center}
\mbox{\epsfig{figure=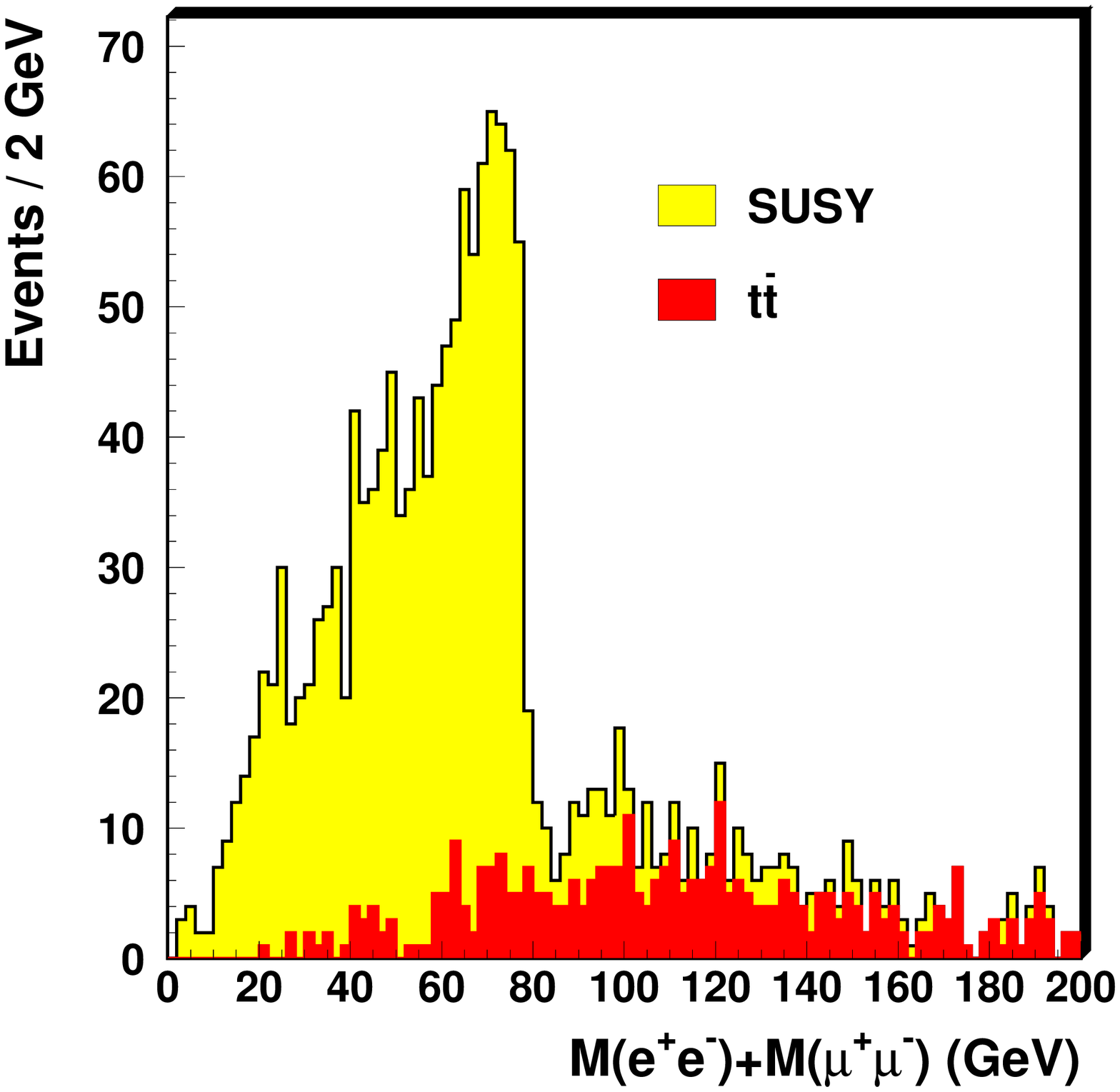,width=0.7\linewidth,clip=}}
\end{center}
\end{minipage}
\caption{(a) Invariant mass distribution of SFOS 
isolated leptons for $\squark$ and $\gluino$ events, superimposed on the
SM background. The contributions of ${\mathrm t\bar{\mathrm t}}$ and Z+jets events are shown. 
(b)Same as in (a) with  ${E_{\mathrm T}^{\mathrm {miss}}} > 150$~GeV cut.\hfill\break }
\label{fig:lepedge}
\end{figure}

Squark and gluino decays are often characterized by long decay chains in which 
the next-to-lightest neutralino ($\Ndue$) is abundantly produced. The $\Ndue$ 
may then decay into the LSP through the chain 
$\Ndue\to\slepton^\pm \ell^\mp\to\Nuno\ell^+\ell^-$. 
Leptons (electrons and muons) from the $\Ndue$ decay exhibit 
a  $\ell^+\ell^-$ invariant mass
distribution with a sharp edge.  If
$m_{\Ndue}<m_{\slepton}+m_\ell$, the $\Ndue$ decay is a three
body decay mediated by a virtual slepton and the edge is placed
at $m_{\Ndue}-m_{\Nuno}$.  Similarly, when
$m_{\Ndue}>m_{\slepton}+m_\ell$, the neutralino decay is a two-body
decay and the corresponding edge is placed at 
\begin{equation}
M_{\ell^+\ell^-} = \frac{\sqrt{ \left( m_{\Ndue}^2 - m_{\slepton}^2 \right) \left( m_{\slepton}^2 
- m_{\Nuno}^2 \right)}}{m_{\slepton}}.
\label{eq:edge}
\end{equation}
Figure~\ref{fig:lepedge} shows the same-flavour opposite-sign (SFOS) dilepton invariant 
mass distribution for a CMS simulation probing the point B of Ref.~\cite{batt}. 
The Standard Model background, due mostly to t${\overline {\mathrm t}}$ and Z+jet events 
(Fig.~\ref{fig:lepedge}a), is rejected by  a cut on the 
transverse missing energy. The remaining contribution is clearly negligible.  
The real background comes from other SUSY processes, and, as is 
illustrated in Fig.~\ref{fig:lepedge}b, is effectively
removed by subtracting the contribution of opposite-sign 
opposite-flavour dilepton pair (OSOF) distribution.  
The edge value for M$_{\ell^+\ell^-}$ is  determined by 
fitting it to a triangular shape with Gaussian smearing. \\
For large values of $\tan\beta$, the $\Ndue$ decays predominantly 
to $\stau$'s (and therefore $\tau$'s). In this configuration, it is more difficult 
to clearly identify the dilepton edge. It is hoped that hadronic $\tau$ 
decays, with appropriate cuts and selections could still lead to an end point in the 
$\tau\tau$ mass distribution, as is shown in Ref.~\cite{atlasph,polnew} 
for the ATLAS Point 6 or for point SPS1A.  

\subsection{Squark and Gluino Mass Reconstruction}
\label{subsec:squark}

\begin{figure}[htb]
\vfill \begin{minipage}{.33\linewidth}
\begin{center}
\mbox{\epsfig{figure=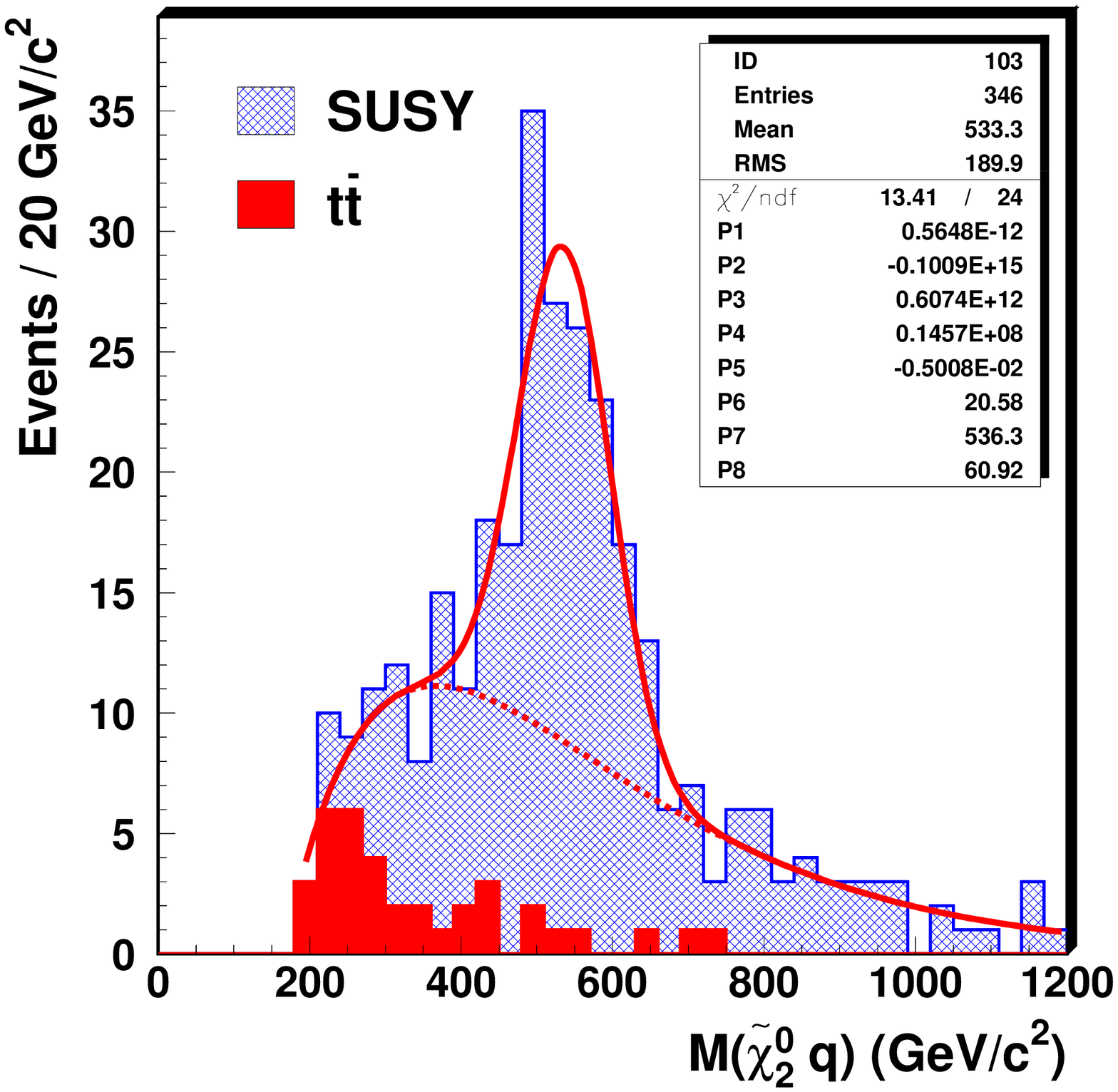,width=0.9\linewidth,clip=}}
\end{center}
\end{minipage}\hfill
\begin{minipage}{.33\linewidth}
\begin{center}
\mbox{\epsfig{figure=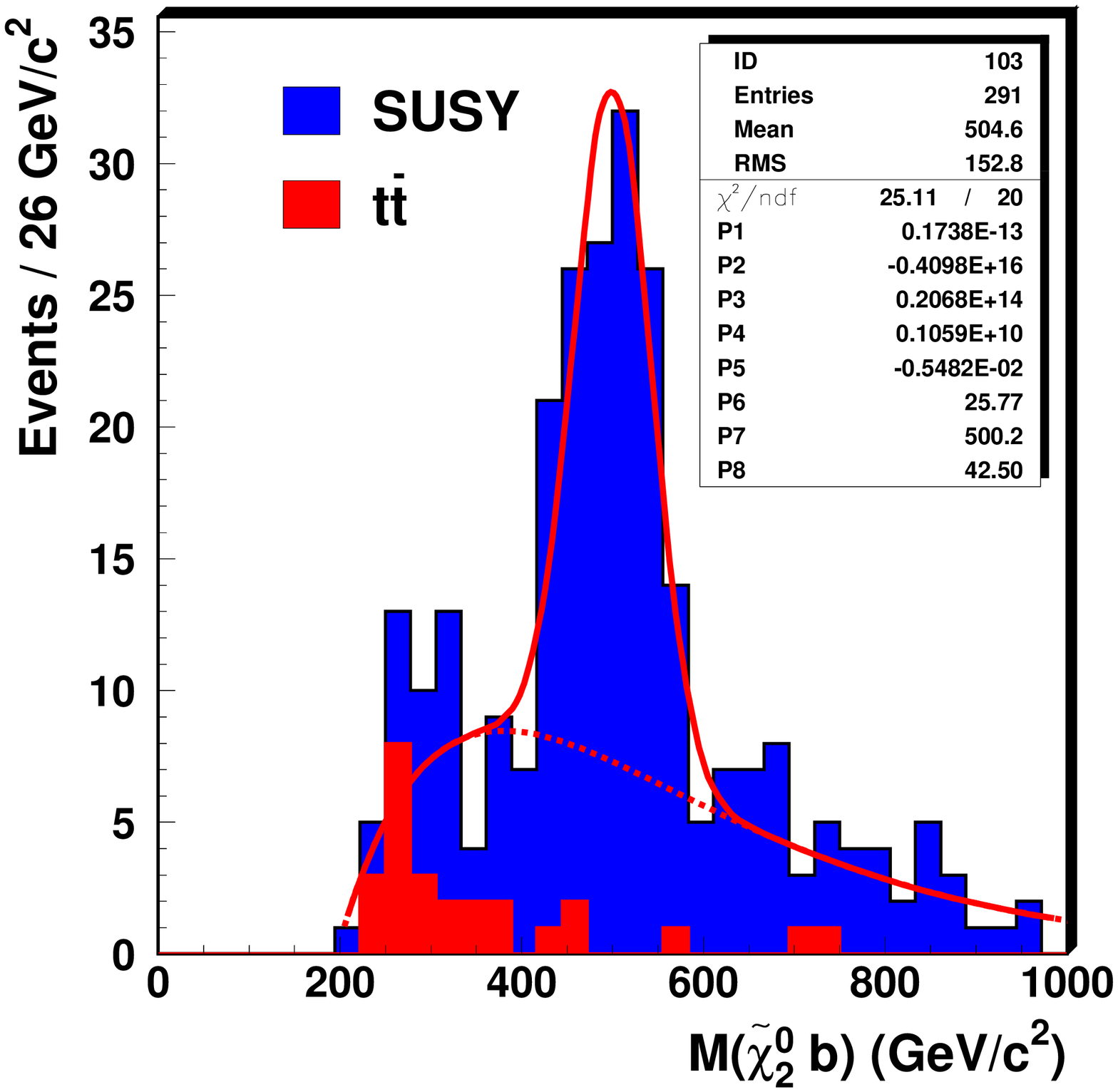,width=0.9\linewidth,clip=}}
\end{center}
\end{minipage}
\begin{minipage}{.33\linewidth}
\begin{center}
\mbox{\epsfig{figure=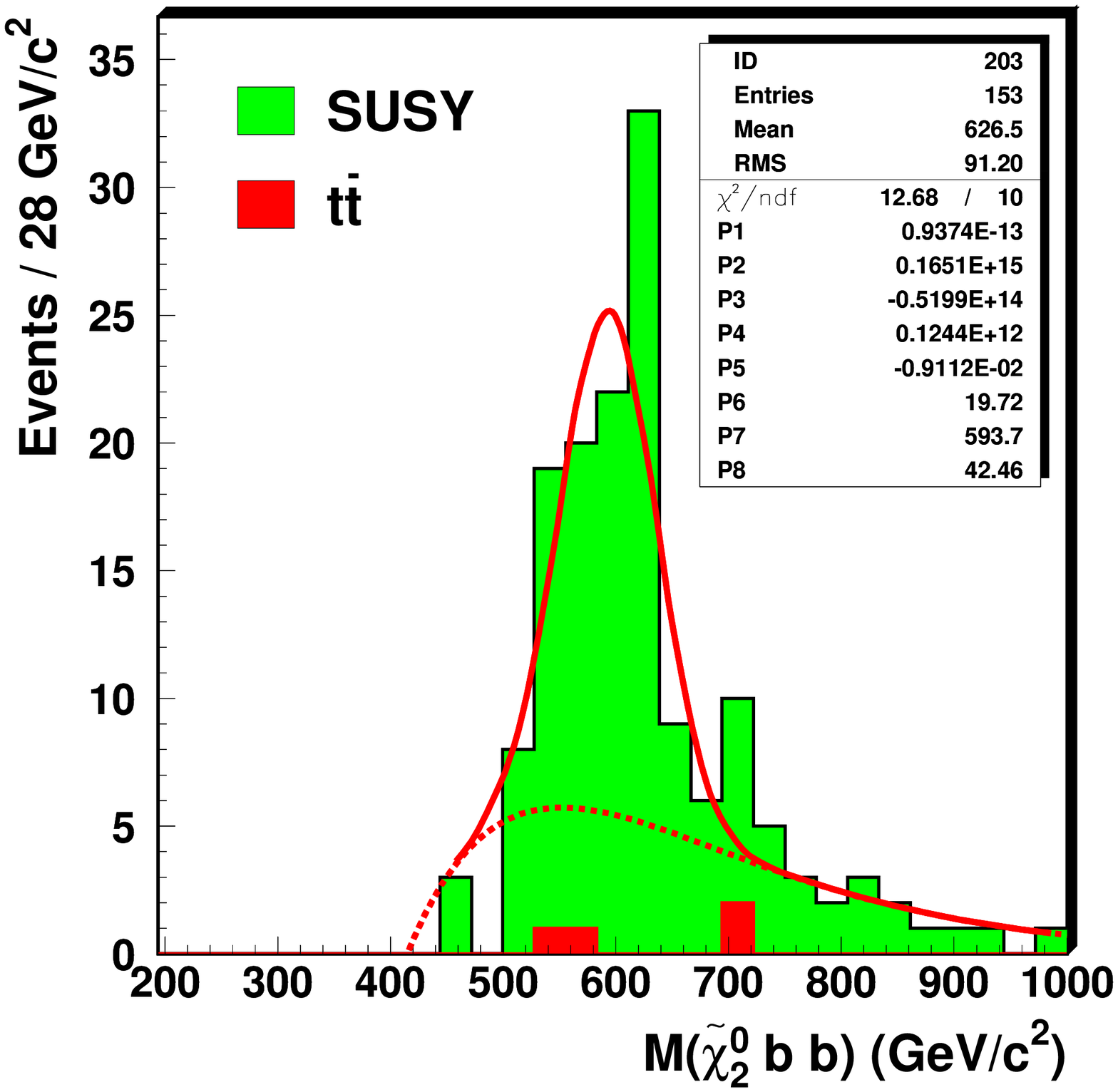,width=0.9\linewidth,clip=}}
\end{center}
\end{minipage}
\caption{Reconstructed invariant mass distribution for squark (left), sbottom (middle) 
and gluino in the sbottom chain (right) at point B. The integrated luminosity is 
1 fb$^{-1}$ for the squark peak and 10 fb$^{-1}$ for the sbottom and the gluino peaks.\hfill\break}
\label{fig:masspeak}
\end{figure}

The reconstruction of a $\Ndue$ is the starting point for more complex decay 
chain reconstruction. Here, the technique used by CMS for squarks and gluino 
mass determination is briefly summarized. Three out of the thirteen ``Post-LEP'' 
benchmark points have been used for this analysis, the point 
B ($m_{1/2}=250$ GeV/$c^2$, $m_0=100$ GeV/$c^2$, $\tan\beta=10$, $\mu>0$ and $A_0=0$), 
G ($m_{1/2}=375$ GeV/$c^2$, $m_0=120$ GeV/$c^2$, $\tan\beta=20$, $\mu>0$ and $A_0=0$) and 
I ($m_{1/2}=350$ GeV/$c^2$, $m_0=180$ GeV/$c^2$, $\tan\beta=35$, $\mu>0$ and $A_0=0$)
of Ref.~\cite{batt}, characterized both by 
relatively low value for $m_0$ and $m_{1/2}$ 
(high $\squark$ and $\gluino$ production cross section) 
and different values of $\tan\beta$. 
In order to perform $\sbottom (\squark)$ and $\gluino$ mass reconstruction,
the decay chain $\gluino\to\sbottom {\mathrm b}, \, \sbottom\to\Ndue {\mathrm b},\,
\Ndue\to\slepton^\pm \ell^\mp\to\Nuno\ell^+\ell^-$, where $\ell={\mathrm e},
\mu$, have been considered. The same decay chain with light quark replacing 
sbottom has been used for squark reconstruction. The events are selected 
by requiring  at least two SFOS isolated leptons
with $p_{\mathrm T} > 15$~GeV/$c$ and $|\eta|<2.4$, corresponding to the acceptance of
the muon system, and two jets (tagged as b jets in the case of sbottom 
reconstruction), with $p_{\mathrm T} >20$~GeV/$c$ and $|\eta|<2.4$. A cut on the missing 
energy is applied in order to suppress the SM background. The first step of 
the reconstruction proceeds as described in the previous paragraph from  
the dilepton edge. 
To reconstruct the sbottom (squark), events in a window 
about 15 GeV/$c^2$ wide around the edge are selected.  This requirement allows 
the kinematical configuration in which the leptons are emitted
back-to-back in the $\Ndue$ rest frame, with the $\Nuno$ at rest, to be selected. In
this configuration the $\Ndue$ momentum is
reconstructed through the relation
\begin{equation}
\vec{p}_{\Ndue} = \left( 1 + \frac{m_{\Nuno}}{M_{\ell^+\ell^-}} \right) \vec{p}_{\ell^+\ell^-}.
\label{eq2}
\end{equation}
Here, the $\Nuno$ mass is assumed to be known. 
A method to extract the $\Nuno$ mass from the data is discussed 
in the next section.
The $\Ndue$ momentum is then added to the momentum of the highest
${\mathrm E_T}$ (b-tagged) jet. The $\squark$ ($\sbottom$) is reconstructed, as shown in 
Figs.~\ref{fig:masspeak}a and ~\ref{fig:masspeak}b. 
To reduce the combinatorial background coming from 
wrong (b) jet associations, further kinematical cuts are used. The reconstruction 
of squarks in a scenario similar to point B is feasible already with an 
integrated luminosity  of 1 fb$^{-1}$, while in the case of the sbottom at 
least 10 fb$^{-1}$ are necessary to achieve a mass resolution 
better than 10\%. Finally, the squark (sbottom) associated with the closest (b) jet in space 
gives the reconstructed gluino mass, as shown in Fig.~\ref{fig:masspeak}c. A resolution 
better than 10\% is achieved also in this case. To summarize, in the case of a precise 
knowledge of the $\Nuno$ mass, the mass resolution achievable for each of the 
sparticles considered is about 10\% for 10 fb$^{-1}$. The dependence of the reconstructed 
masses on the $\Nuno$ mass uncertainty  has been evaluated in Ref.~\cite{noi} to be  
\bea
\Delta m(\Ndue {\mathrm b}) = (1.60\pm 0.03)\Delta m(\Nuno).
\eea
  
Although  both $m$($\squark$) and $m$($\gluino$) depend on the $\Nuno$ mass, their difference 
is in contrast independent of m($\Nuno$) and can be measured with an error of few 
percent, irrespective of the sparticle spectrum~\cite{noi}. 

All the results shown so far are derived for point B (with parameters 
similar to that of point SPS1A) and for an
integrated luminosity of 10~fb$^{-1}$.  The same kind of analysis was
repeated also for point G. In this case, however, 
the higher value of $\tan\beta$ reflects into higher
branching ratio for the decay $\Ndue \rightarrow \tau^+\tau^- \Nuno$, 
hence  to a smaller signal $\Ndue\to\slepton^\pm
\ell^\mp\to\Nuno\ell^+\ell^-, \,\ell={\mathrm e},\mu$. Reconstruction of squarks 
and gluinos is only possible with tighter cuts and with high 
integrated luminosity. An attempt was made  to repeat 
the analysis at point I of Ref.~\cite{batt}, characterized by 
a yet higher value of $\tan\beta$ ($\tan\beta=35$). For that point, 
even with an integrated luminosity of 300~fb$^{-1}$, it is not possible 
to reconstruct squarks and gluinos with this method. 

\subsection{$\Nuno$ Mass Determination}
\label{subsec:nuno}

\begin{figure}[h!]
\begin{center}
\mbox{\epsfig{figure=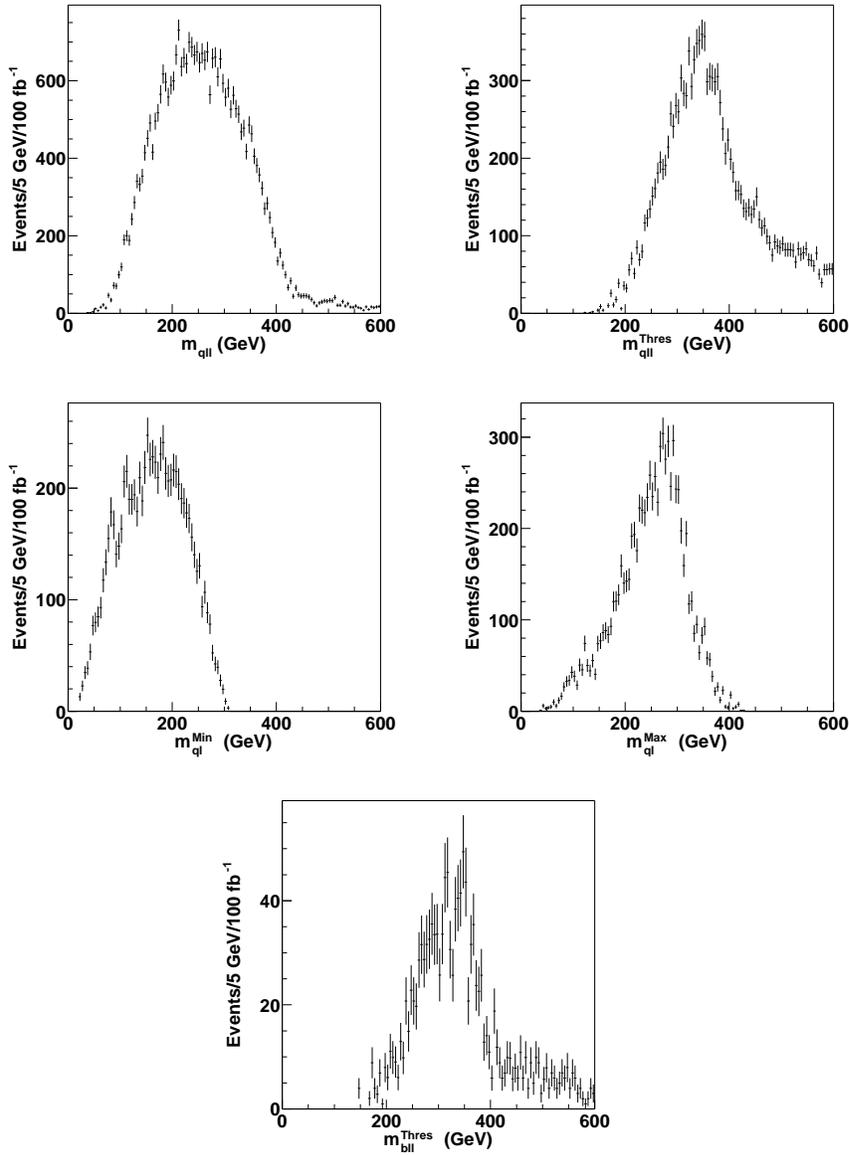,height=1.0\linewidth,clip=}}
\end{center}
\caption{ATLAS invariant mass distributions with kinematical end points for the mSUGRA 
Point SPS1A.}
\label{fig:endpoints}
\end{figure}

\begin{figure}[htb]
\vfill \begin{minipage}{.49\linewidth}
\begin{center}
\mbox{\epsfig{figure=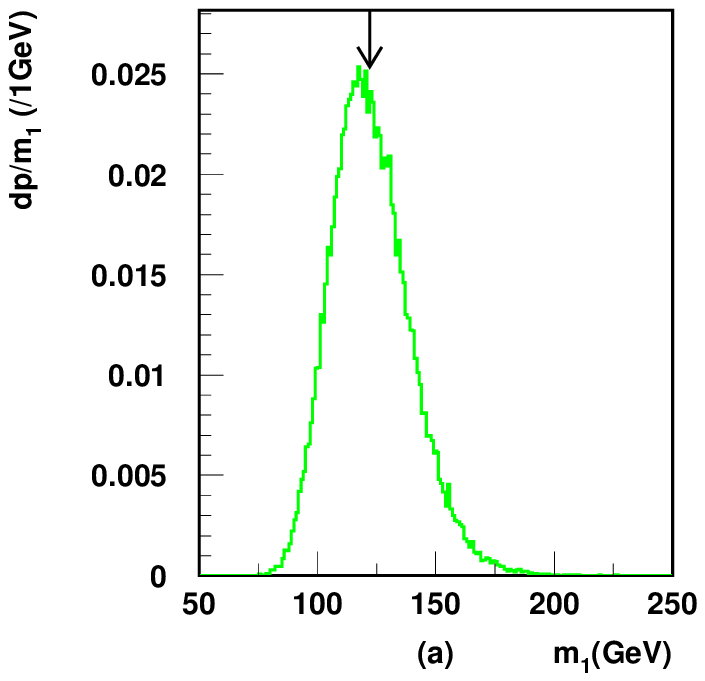,height=0.85\linewidth,clip=}}
\end{center}
\caption{$M$($\Nuno$) distribution for ATLAS point 5.}
\label{fig:nuno}
\end{minipage}\hfill
\begin{minipage}{.49\linewidth}
\begin{center}
\mbox{\epsfig{figure=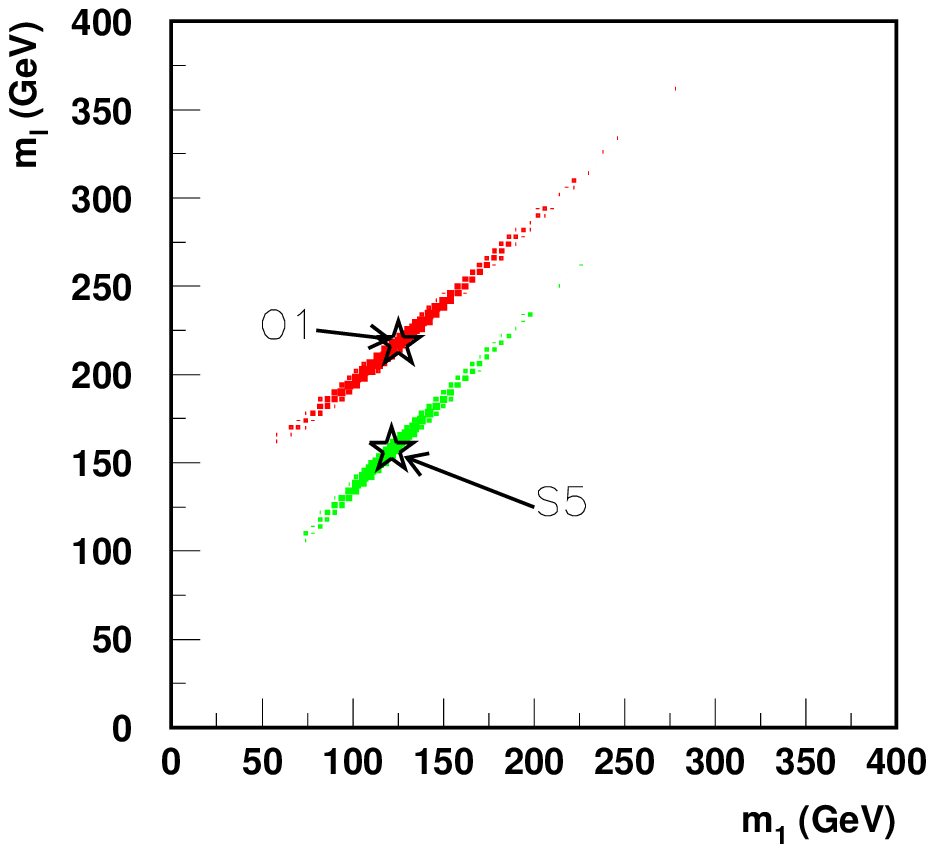,width=0.95\linewidth,clip=}}
\end{center}
\caption{Reconstructed $M$($\slepton$) vs $M$($\Nuno$) for O1 and S5 (see text). 
The stars show the true mass for each model.} 
\label{fig:ben}
\end{minipage}
\end{figure}

In sequential decays, the presence of other end points and thresholds, 
as suggested in Ref.~\cite{allanach}, can be used to extract, in a 
model-independent way, the masses of the sparticles, and, finally,  
to disentangle different models. The starting point is the same 
decay chain as in the previous section. The two leading jets are assumed to 
come from the squark decays and are combined with the lepton to 
find other end points and thresholds. For example, the largest dilepton-jet mass,  
$M_{\ell\ell {\mathrm q}}$ and the largest lepton-jet masses, for the first and the second 
lepton, $M_{\ell {\mathrm q}}^{\mathrm {max}}$ and 
$M_{\ell {\mathrm q}}^{\mathrm {min}}$, are expected to give end points as well as  
$M_{\ell\ell {\mathrm q}}$
should give a threshold. The distributions of these 
quantities are shown in Fig.~\ref{fig:endpoints}. Other end points 
are clearly visible and well measurable, despite various detection and 
reconstruction spoiling factors. With the high statistics reachable at LHC, it seems 
possible to measure the end points with a precision limited only by the hadronic 
scale accuracy (1 to 2\% depending on the particle combinations for point B or SPS1A). \\
It is possibile to extract the masses of the sparticles with a combination of all these 
end point measurements. In particular, 
the mass of $\Nuno$ is a fundamental ingredient of the other sparticle mass 
measurements. 
In Fig.~\ref{fig:nuno}, the reconstructed $\Nuno$ mass peak is shown. A resolution 
of about 10\% is achievable, within a given model. 

In Fig.~\ref{fig:ben} the result of the fits with the mSUGRA Point 5 (S5) and the point in 
an Optimized String Model (O1)~\cite{allanach} are 
compared. 

\section*{Conclusions}

If SUSY exists at the EW scale, both the ATLAS and CMS will be able to 
discover it over a large range of the paramater space. Squark and
gluino decays present characteristic signatures to discriminate the
SUSY processes from the Standard Model. Inclusive studies have
demonstrated that squarks and gluinos could be discovered already in
the first months of data taking. With the ultimate luminosity of
300~fb$^{-1}$, strongly interacting sparticles could be discovered up
to masses of 2.5 to 3 TeV/$c^2$. 

\noindent Altough sparticle reconstruction is more difficult, new analyses have
shown that in some cases it is possible to make exclusive
reconstructions. A search for special features, like kinematical 
end points and thresholds, excesses of b/$\tau$'s, isolated leptons  
helps in the reconstruction of full decay chains and, possibly, 
to disentangle different theoretical models. 
This is the case, for instance, for the decay
$\gluino\to\squark (\sbottom) {\mathrm q}({\mathrm b})$ which allows  both 
squark (sbottom) and gluino masses to be reconstructed. Resolutions better 
than 10\% are attainable in the
low $\tan\beta$ region, already after the first year of data
taking. More work is in progress to evaluate the ATLAS and CMS capability to
reconstruct SUSY sparticles.

\section*{Acknowledgments}
I would like to thank the Conference Organizers for the invitation and for the friendly 
hospitality. Special thanks also to M. Chiorboli and G. Polesello for providing materials 
for the presentation and helpful discussion. I would like also to thank P. Janot for the
rewieving.

\end{document}